\documentclass[twocolumn,prl,floats,showpacs]{revtex4}
\usepackage{graphics}
\usepackage{amsmath}
\usepackage{graphicx}
\usepackage{amsfonts}
\usepackage{amssymb}

\begin{document}
\title{Matter-wave amplification and phase conjugation via stimulated dissociation of a molecular Bose-Einstein condensate }
\author{Kar\'{e}n V. Kheruntsyan}
\affiliation{ARC Centre of Excellence for Quantum-Atom Optics, Department of Physics,
University of Queensland, Brisbane, Queensland 4072, Australia}
\date{\today}

\begin{abstract}
We propose a scheme for parametric amplification and
phase conjugation of an atomic Bose-Einstein condensate (BEC) via stimulated
dissociation of a BEC of molecular dimers consisting of bosonic atoms. This
can potentially be realized via coherent Raman transitions or using a magnetic
Feshbach resonance. We show that the interaction of a small incoming atomic BEC
with a (stationary) molecular BEC can produce two
counterpropagating atomic beams -- an amplified atomic BEC and its
phase-conjugate or \textquotedblleft time-reversed\textquotedblright\ replica.
The two beams can possess strong quantum correlation in the relative particle
number, with squeezed number-difference fluctuations.

\end{abstract}

\maketitle

The fascinating experimental progress in controlling ultracold atomic gases
has now resulted in the creation of quantum-degenerate samples of ultracold
molecules and molecular Bose-Einstein condensates (BEC). The first step
toward seeing molecular condensation was undertaken in transient experiments
with a BEC of $^{85}$Rb atoms \cite{Donley}, in which interference
measurements were indicative of a small molecular condensate formation. More
recent experiments with $^{133}$Cs, $^{87}$Rb, and $^{23}$Na
\cite{Bosonic-Cs-Rb-Na}, as well as with degenerate Fermi gases of $^{40}$K
and $^{6}$Li atoms \cite{Fermionic}, have produced even larger samples of
ultracold molecules, including molecular condensates.

All these experiments have employed conversion of atom pairs into weakly bound
molecular dimers in the vicinity of a magnetically tunable Feshbach resonance.
This technique appears to be more successful at present than Raman
photoassociation \cite{Heinzen-exp,Superchemistry}. Unambiguous claims on the
observation of equilibrium molecular condensates have only been made for the
case of molecules consisting of fermionic atoms. These molecules have much
longer lifetimes because of Pauli blocking, which suppresses their decay due to
inelastic collisions \cite{Petrov}. Despite this, the production of Bose
condensed dimers composed of bosonic atoms should not be out of reach using
current experimental techniques.

The purely bosonic case is particularly interesting since the underlying
dynamics in the atom-molecule conversion can take advantage of coherence and
bosonic stimulation with respect to both the atomic and molecular species. In
practical terms, this could lead to wider range of techniques for coherent quantum
control of ultracold quantum gases, including the possibility of
superchemistry \cite{Superchemistry}, matter-wave amplification with
atom-molecular \textquotedblleft laser\textquotedblright\ beams, as well as
various implementations of nonlinear atom optics and quantum atoms optics.

Here, we propose a scheme for nonlinear, coherent matter-wave amplification
and phase conjugation, which at the same time is capable of producing
pair-correlated or number squeezed atomic beams. The scheme relies on the
process of stimulated dissociation of a BEC of molecular dimers, taking place
in the presence of an injected signal -- a small incoming atomic condensate.
The resulting output consists of an amplified signal and its \textquotedblleft
time-reversed\textquotedblright\ or phase-conjugate replica \cite{Meystre}.

This is analogous to the nonlinear optical process of parametric amplification
of a \textquotedblleft seed\textquotedblright\ signal, using frequency
conversion of photons in a quadratically nonlinear media \cite{Yariv}. In the
matter-wave parametric amplification, the coupling that takes the role of the
\textquotedblleft quadratic nonlinearity\textquotedblright\ and converts
molecules into atom pairs and vice versa can be realized via coherent Raman
transitions or using magnetic Feshbach resonances. Either of these mechanisms
have certain advantages and disadvantages from the practical point of view,
however, the essential physics can be modeled via an effective quantum-field 
theory, which is identical in the two cases \cite{PDKKHH-1998,AM-BEC via
PA,AM-BEC via Feshbach}.

The advantage of the present scheme compared to spontaneous
molecule dissociation
\cite{Moelmer2001,TwinBeams,Yurovsky,Dissociation-exp,Rempe}, with no
atomic condensate present initially, is that the desired quantum
effects with mesoscopic atom numbers can be achieved on much
shorter time scales. As a result, the disruptive effect of
inelastic collisions on molecule lifetimes can be bypassed, thus
making the present scheme much more robust for practical
implementation. In addition, the scheme has an advantage compared
to earlier related proposals \cite{4WM-theory} using four-wave
mixing \cite{4WM-exp} in that it is less susceptible to phase
noise in the \textquotedblleft pump\textquotedblright\ BEC -- once
its depletion is taken into account.

We start by considering the Heisenberg equations of motion for the coupled
atomic-molecular system in a one-dimensional (1D) environment, which in a
rotating frame and in a dimensionless form are given by
\cite{TwinBeams,PDKKHH-1998}
\begin{align}
\frac{\partial\hat{\psi}_{1}}{\partial\tau}  &  =i\frac{\partial^{2}\hat{\psi
}_{1}}{\partial\xi^{2}}-i\delta\hat{\psi}_{1}+\kappa\hat{\psi}_{2}\hat{\psi
}_{1}^{\dag},\nonumber\\
\frac{\partial\hat{\psi}_{2}}{\partial\tau}  &  =\frac{i}{2}\frac{\partial
^{2}\hat{\psi}_{2}}{\partial\xi^{2}}-i\left[  V_{2}(\xi)+u_{22}\hat{\psi}
_{2}^{\dagger}\hat{\psi}_{2}\right]  \hat{\psi}_{2}-\frac{1}{2}\kappa\hat
{\psi}_{1}^{2}. \label{Heizenberg-Eqs}
\end{align}
Here, $\hat{\psi}_{1}(\xi,\tau)=\sqrt{d_{0}}\hat{\Psi}_{1}(x,t)$ and
$\hat{\psi}_{2}(\xi,\tau)=\sqrt{d_{0}}\hat{\Psi}_{2}(x,t)$ represent the
dimensionless atomic and molecular field operators (such that $\left\langle
\hat{\Psi}_{i}^{\dagger}\hat{\Psi}_{i}\right\rangle $ gives the 1D linear
density), with bosonic commutation relations $[\hat{\psi}_{i}(\xi,\tau
),\hat{\psi}_{j}^{\dagger}(\xi^{\prime},\tau)]=\delta_{ij}\delta(\xi
-\xi^{\prime})$, where $\xi=x/d_{0}$ is the dimensionless coordinate,
$\tau=t/t_{0}$ is the dimensionless time, $d_{0}$ is a length scale, and
$t_{0}=2m_{1}d_{0}^{2}/\hbar$ is a time scale. In addition, $m_{1}$ and
$m_{2}=2m_{1}$ are the atomic and molecular masses, $\delta=\Delta t_{0}$ is
the dimensionless detuning (with $2\hbar\Delta$ giving the overall energy
mismatch between a pair of free atoms and a molecule), $\kappa=\chi
t_{0}/\sqrt{d}_{0}$ is the dimensionless atom-molecule coupling (with $\chi$
being the respective 1D coupling), and $u_{22}=U_{22}t_{0}/d_{0}$ is the
dimensionless coupling for molecule-molecule $s$-wave scattering interaction
(with $U_{22}$ being the respective 1D coupling
\cite{Olshanii98,KK-DG-PD-GS-2003}). Finally, $V_{2}(\xi)=-u_{22}n_{2}
^{0}(1-\xi^{2}/\xi_{0}^{2})$ describes the axial trapping potential for the
molecules which we assume is harmonic, where $n_{2}^{0}=\left\langle \hat
{\psi}_{2}^{\dagger}(0,0)\hat{\psi}_{2}(0,0)\right\rangle $ is the
dimensionless initial peak density of the molecular condensate and $\xi_{0}$
is the respective Thomas-Fermi radius.

We assume a highly elongated cigar-shaped trap geometry such that the system
can be modelled by a 1D theory \cite{Olshanii98,KK-DG-PD-GS-2003}, as above.
The atoms are assumed to be untrapped axially, while confined transversely.
The molecules are trapped both transversely and axially.

In the case of Raman photodissociation, the energy mismatch is given by
$2\hbar\Delta=2E_{1}-E_{2}-\hbar\omega$, where $\omega$ is the frequency
difference between the two Raman lasers, while $2E_{1}$ and $E_{2}$ refer,
respectively, to the energy of free-atom pairs in the dissociation limit and
the energy of the bound molecular state.

In the case of a Feshbach resonance \cite{Dissociation-exp,Rempe}, $2\hbar\Delta$
gives the energy mismatch achieved upon switching on the atom-molecule
coupling, i.e., upon a rapid crossing of the magnetic field through the
resonance, which \textquotedblleft brings\textquotedblright\ the (initially
stable) molecular level above the atomic dissociation limit. This corresponds to
having $\Delta<0$, and results in dissociation of molecules into atom pairs
such that the potential energy $2\hbar|\Delta|$ is converted into the atomic
kinetic energy [$2\hbar|\Delta|\simeq2\hbar^{2}k^{2}/(2m_{1})$] for selected
phase-matched modes with opposite momenta around $\pm k_{0}$ $=\sqrt
{2m_{1}|\Delta|/\hbar}$. This strategy of a fast ramp through the resonance 
is very similar to the one realized experimentally in Ref. \cite{Rempe}, where 
a quasi-mono-energetic spherical wave of atoms was created. A realization of 
this experiment in a one-dimensional environment would correspond to the 
conditions of the current proposal.

In our model, the interaction terms due to atom-atom and atom-molecule $s$-wave 
scattering processes are neglected on the grounds that we only consider large absolute values
of the detuning $|\Delta|$ and restrict ourselves to short dissociation times.
As a result, the number of atoms produced during dissociation remains small,
and the mean-field or phase diffusion terms due to atom-atom and atom-molecule
scatterings remain negligible compared to $|\Delta|$.

Before analyzing the quantum dynamics of the system described by Eqs.
(\ref{Heizenberg-Eqs}), we first consider a simplified model that has
an analytic solution. This corresponds to an undepleted, uniform molecular
condensate in a coherent state at density $n_{2}^{0}$, in which case the
molecular field amplitude (which we assume is real) can be absorbed into an
effective coupling $g=\kappa\sqrt{n_{2}^{0}}$. Expanding $\hat{\psi}_{1}
(\xi,\tau)$ in terms of single-mode annihilation operators, $\hat{\psi}
_{1}(\xi,\tau)=\sum_{q}\hat{a}_{q}(\tau)e^{iq\xi}/\sqrt{l}$, where $q=kd_{0}$
is the dimensionless momentum, $l$ is the quantization length, and the
operators $\hat{a}_{q}$ satisfy the usual commutation relations $[\hat{a}
_{q}(\tau),\hat{a}_{q^{\prime}}^{\dagger}(\tau)]=\delta_{q,q^{\prime}}$, we
obtain the following Heisenberg equations of motion:
\begin{align}
d\hat{a}_{q}/d\tau &  =-i\left[  q^{2}+\delta\right]
\hat{a}_{q}+g\hat
{a}_{-q}^{\dagger},\nonumber\\
d\hat{a}_{-q}^{\dagger}/d\tau &  =i\left[  q^{2}+\delta\right]
\hat{a} _{-q}^{\dagger}+g\hat{a}_{q}. \label{a_k_equations}
\end{align}

These have the following well-known solution \cite{TwinBeams,Yurke}: $\hat
{a}_{q}(\tau)=A_{q}(\tau)\hat{a}_{q}(0)+B_{q}(\tau)\hat{a}_{-q}^{\dagger}(0)$
and $\hat{a}_{-q}^{\dagger}(\tau)=B_{q}(\tau)\hat{a}_{q}(0)+A_{q}^{\ast}
(\tau)\hat{a}_{-q}^{\dagger}(0)$, where
\begin{align}
A_{q}(\tau)  &  =\cosh\left(  g_{q}\tau\right)  -i\lambda_{q}\sinh\left(
g_{q}\tau\right)  /g_{q},\nonumber\\
B_{q}(\tau)  &  =g\sinh\left(  g_{q}\tau\right)  /g_{q},
\end{align}
and $|A_{q}|^{2}-B_{q}^{2}=1$. Here, the parameter $\lambda_{q}\equiv
q^{2}+\delta$ can be identified with an \emph{effective} phase mismatch, while
$g_{q}\equiv(g^{2}-\lambda_{q}^{2})^{1/2}$ is the gain coefficient, which --
if real -- causes a growing output in mode $q$, while -- if imaginary -- leads
to oscillations.

To study parametric amplification in the atomic field, we consider initial
conditions where all momentum components with $q>0$ are initially in a
coherent state $\hat{a}_{q}(0)\left\vert \alpha_{q}\right\rangle =\alpha
_{q}\left\vert \alpha_{q}\right\rangle $, where $\alpha_{q}$ are the
corresponding amplitudes, while all negative momentum components are initially
in the vacuum state $\hat{a}_{-q}(0)\left\vert 0\right\rangle =0$.

Next, we introduce particle number operators $\hat{N}_{+}=\sum\nolimits_{q>0}
\hat{a}_{q}^{\dagger}\hat{a}_{q}$ and $\hat{N}_{-}=\sum\nolimits_{q>0}\hat
{a}_{-q}^{\dagger}\hat{a}_{-q}$ corresponding to the total number of atoms
with positive and negative momenta, respectively, and find that the respective
average numbers are given by:
\begin{align}
\langle\hat{N}_{+}(\tau)\rangle &  =\sum\nolimits_{q>0}\left[  B_{q}^{2}
(\tau)\left(  1+|\alpha_{q}|^{2}\right)  +|\alpha_{q}|^{2}\right]  ,\\
\langle\hat{N}_{-}(\tau)\rangle &  =\sum\nolimits_{q>0}B_{q}^{2}
(\tau)(1+|\alpha_{q}|^{2}).
\end{align}
For $\delta<0$, the function $B_{q}^{2}$ has two distinct global maxima at
$\pm q_{0}=\sqrt{|\delta|}$ corresponding to a zero effective phase mismatch,
$\lambda_{q}=0$. In the expression for $\langle\hat{N}_{+}(\tau)\rangle$, the
three terms under the sum are identified as the amplified contribution of the
vacuum noise in the mode $q$, $B_{q}^{2}(\tau)$, the amplified coherent
component of the input, $B_{q}^{2}(\tau)|\alpha_{q}|^{2}$, and the coherent
input component itself, $|\alpha_{q}|^{2}$, while $\langle\hat{N}_{-}
(\tau)\rangle$ consists of the amplified vacuum noise and the phase conjugate
of the input.

To quantify the correlation and relative number squeezing between
$\hat{N}_{+}$ and $\hat{N}_{-}$, we consider the normalized
variance $V(\tau)$  of the particle number difference
$\hat{N}_{+}(\tau)-\hat{N} _{-}(\tau)$. In normally ordered form,
$V(\tau)$ is given by
\begin{equation}
V(\tau)=1+\langle:[\Delta(\hat{N}_{+}-\hat{N}_{-})]^{2}:\rangle/(\langle
\hat{N}_{+}\rangle+\langle\hat{N}_{-}\rangle). \label{V-define}
\end{equation}
Here, $\Delta\hat{X}\equiv\hat{X}-\langle\hat{X}\rangle$, and
$V(\tau)<1$ implies squeezing of fluctuations below the coherent
level, which is due to strong quantum correlation between the
particle numbers in $\hat{N}_{+}(\tau)$ and $\hat{N}_{-}(\tau)$.

Calculating the quantities $\langle:(\hat{N}_{+,-})^{2}:\rangle$
and $\langle\hat{N}_{+}\hat{N}_{-}\rangle$ in
Eq.~(\ref{V-define}) and assuming that $|\alpha_{q}|^{2}\gg1$
gives the following approximate result for the variance:
\begin{equation}
V(\tau)\simeq1-\frac{2\sum\nolimits_{q>0}B_{q}^{2}(\tau)|\alpha_{q}|^{2}}
{\sum\nolimits_{q>0}[2B_{q}^{2}(\tau)+1]|\alpha_{q}|^{2}}. \label{V-approx}
\end{equation}
As we see, the degree of squeezing depends on the magnitude of the
amplification factor $B_{q}^{2}(\tau)$, and for strong amplification,
$B_{q}^{2}(\tau)\gg1$, one can obtain almost perfect ($100\%$) squeezing,
$V(\tau)\simeq0$.

We now turn to the exact quantum dynamical simulation of the
\emph{nonuniform} system, Eqs. (\ref{Heizenberg-Eqs}). Here, we
take into account molecular field depletion, molecule-molecule
$s$-wave scattering, and we include possible (linear) losses of
atoms and molecules, occurring at rates $\gamma_{1}$ and
$\gamma_{2}$, respectively. The simulation is done via numerical
solution of the stochastic ($c$-number) differential equations
\cite{TwinBeams} in the positive-$P$ representation \cite{+P}
\begin{align}
\frac{\partial\psi_{1}}{\partial\tau}  &  =i\frac{\partial^{2}\psi_{1}
}{\partial\xi^{2}}-(\gamma_{1}+i\delta)\psi_{1}+\kappa\psi_{2}\psi_{1}
^{+}+\sqrt{\kappa\psi_{2}}\eta_{1},\nonumber\\
\frac{\partial\psi_{2}}{\partial\tau}  &  =\frac{i}{2}\frac{\partial^{2}
\psi_{2}}{\partial\xi^{2}}-\left[  \gamma_{2}+iV_{2}(\xi)+iu_{22}\psi_{2}
^{+}\psi_{2}\right]  \psi_{2}\nonumber\\
& -\frac{\kappa}{2}\psi_{1}^{2}+\sqrt{-iu_{22}}\psi_{2}\eta_{2},
\label{Positive-P-eqs}
\end{align}
together with the equations for the \textquotedblleft
conjugate\textquotedblright\ fields $\psi_{1,2}^{+}$, having noise
terms $\eta_{1,2}^{+}$. Apart from the new loss terms, these
equations are equivalent to Eqs. (\ref{Heizenberg-Eqs}), where
$\psi_{i}$ and $\psi_{i}^{+}$ are independent complex stochastic
fields corresponding, respectively, to the operators
$\hat{\psi}_{i}$ and $\hat{\psi}_{i}^{\dag}$, while $\eta_{i}$ and
$\eta_{i}^{+}$ ($i=1,2$) are four real independent
$\delta$-correlated Gaussian noises with $\left\langle
\eta_{i}(\xi,\tau)\eta_{j}(\xi^{\prime},\tau
^{\prime})\right\rangle
=\langle\eta_{i}^{+}(\xi,\tau)\eta_{j}^{+}(\xi
^{\prime},\tau^{\prime})\rangle=\delta_{ij}\delta(\xi-\xi^{\prime})\delta
(\tau-\tau^{\prime})$.
\begin{figure}[ptb]
\centering
\includegraphics[height=4.5cm]{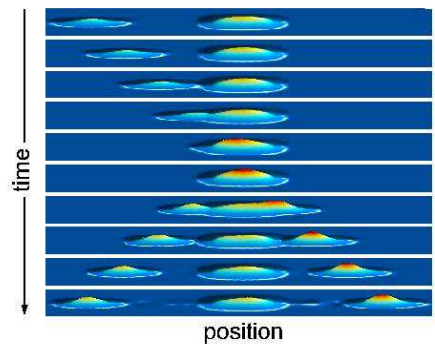}
\caption{Atomic and molecular (shown in the middle) density
profiles, illustrating parametric amplification and
phase conjugation of an incoming atomic BEC (top frame, left) via
stimulated molecule dissociation \cite{Figure1-comment}. The
results are obtained using 5,000 stochastic trajectory averages
for $\kappa=84$, $u_{22}=1.8$, $\delta=-4.9\times10^{4}$,
$\gamma_{1}=0.5$, and $\gamma_{2}=0$. The dissociation coupling is
switched on at $\tau_{1}=0.01$ for a duration of
$\Delta\tau=4\times10^{-4}$, while the total time window is
$\tau_{f}=0.0204$.} \label{Fig1}
\end{figure}

Figure \ref{Fig1} represents an example of a quantum dynamical simulation of
Eqs.~(\ref{Positive-P-eqs}) illustrating parametric amplification and
phase conjugation of an incident atomic BEC with a center-of-mass momentum
$q_{0}=\sqrt{|\delta|}$. Here, the top frame corresponds to the initial
condition of a stable molecular condensate (shown in the middle) in a coherent
state with a Thomas-Fermi density profile. The small incoming atomic BEC
(shown on the left and moving to the right) is also assumed to be in a
coherent state initially, and since we neglect the atom-atom interactions in
this low-density regime, we assume a Gaussian density profile. The
dissociation coupling $\kappa$ is invoked at time $\tau_{1}$ when the atomic
cloud is aligned with the molecular BEC (frame 5 from top). It is kept
switched on for a short duration $\Delta\tau$ such that the amplified and the
\textquotedblleft reflected\textquotedblright\ output beams have densities
comparable to that of the input beam and can be seen on the same graph.

The next set of simulations is carried out for a more realistic
set of parameter values than in Fig.~\ref{Fig1}. We use a longer
dissociation time and a larger coupling $\kappa$ -- to result in a
large amplification factor and hence a strongly correlated output
with squeezing in the particle number difference. For simplicity,
the simulation starts from $\tau=\tau_{1}=0$ when the incoming
atomic cloud is already aligned with the molecular condensate and
we invoke the atom-molecule coupling $\kappa$. The dissociation is
then stopped ($\kappa=0$) at $\tau =\tau_{2}$, and we continue the
dynamical evolution of the atomic field in free space (in 1D) to
allow spatial separation of the modes with positive and negative
momenta during the time interval from $\tau_{2}$ to $\tau_{f}$.

For spatially separated components, we can introduce a pair of particle number
operators
\begin{equation}
\hat{N}_{+(-)}(\tau)=\vspace{-0.25cm}\int_{0(-l/2)}^{l/2(0)}\hat{\psi}
_{1}^{\dag}(\xi,\tau)\hat{\psi}_{1}(\xi,\tau)d\xi.
\end{equation}
Next, we define the normalized variance $V(\tau)$ of the particle number
difference as in Eq. (\ref{V-define}) and evaluate numerically the relevant 
averages, using the standard correspondence between the normally-ordered
operator moments and the $c$-number stochastic averages \cite{+P}.
\begin{figure}[ptb]
\centering
\includegraphics[width=7cm]{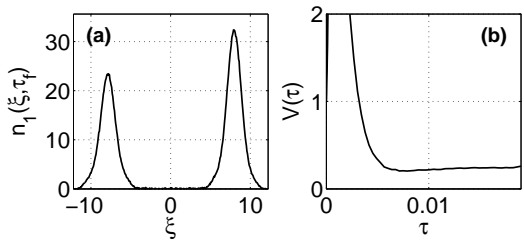}
\caption{(a) Final atomic density profile $n_{1}(\protect\xi ,\protect\tau %
_{f})$ and (b) the variance $V(\protect\tau )$ as a function of time $%
\protect\tau $. Here, the simulation (with 40,000 stochastic trajectory
averages) starts at $\protect\tau _{1}$, the duration of dissociation is $%
\Delta \protect\tau =8\times 10^{-4}$, and $\protect\tau _{f}=1.9\times
10^{-2}$. Other parameter values are $\protect\kappa =297$, $u_{22}=0.068$, 
$\protect\delta =-5.52\times 10^{4}$, $\protect\gamma _{1}=$ $\protect%
\gamma _{2}=3$, $n_{2}^{0}=89.3$, and $\protect\xi _{0}=3.49$ \protect\cite%
{Figure2-comment}.}
\label{Fig2}
\end{figure}

Figure \ref{Fig2} shows the results for the atomic density
distribution
$n_{1}(\xi,\tau_{f})=\langle\hat{\psi}_{1}^{\dag}(\xi,\tau_{f})\hat{\psi}
_{1}(\xi,\tau_{f})\rangle$ at final time $\tau=\tau_{f}$ and the
variance $V(\tau)$ as a function of $\tau$. In this simulation,
the initial total number of molecules is $415$, the initial number
of atoms in the incoming BEC is $21$, whereas the final 
number of atoms in the two output beams is $\sim60$ and $78$. The
squeezing in the particle number difference at $\tau=\tau_{f}$ is
about $75\%$ [$V(\tau_{f})\simeq0.25$], and it is achieved on a
time scale of $\Delta\tau=\tau_{2}-\tau_{1}=8\times10^{-4}$
corresponding to $12.7$ ms, with the parameter values used here
\cite{Figure2-comment}. This is much shorter than the time scale
required in spontaneous dissociation \cite{TwinBeams} to reach the
same total number of atoms. At the same time, the degree of
squeezing is still rather high.

Thus, the disruptive effect of molecule losses due to inelastic collisions can
be reduced in the present scheme. The reason for the shorter time scales
required here is that the process of dissociation in the presence of a
\textquotedblleft seed\textquotedblright\ atomic BEC begins in the stimulated
regime, with exponentially growing output. In the case of spontaneous
dissociation, on the other hand, the initial dynamics is in the spontaneous
regime and the system spends a relatively long time here before the bosonic
stimulation into the phase-matched modes becomes dominant.

To summarize, we have analyzed the process of stimulated
dissociation of a condensate of molecular dimers in the presence
of a small incoming atomic BEC. This results in parametric
amplification of the input BEC together with generation of its
phase-conjugate replica, propagating in the opposite direction.
The two output beams are strongly correlated in the particle
number and have squeezed number-difference fluctuations. The
squeezing with a mesoscopic total number of atoms can be achieved
on much shorter time scales than in the case of spontaneous
dissociation. This makes the present scheme more feasible for
practical implementation, using short-lived molecular condensates.
In addition, the scheme provides a range of opportunities for coherent
quantum control of ultracold quantum gases, including 
applications of nonlinear and quantum atom optics.

The author thanks P. D. Drummond and M. Olsen for helpful
discussions, and the authors of the XmdS software \cite{xmds} used
for simulations. This work was supported by the Australian
Research Council and by the National Science Foundation under
Grant No. PHY99-07949.


\begin{thebibliography}{99}
\bibitem {Figure1-comment}For illustration purposes, we use different scales
for the atomic and molecular densities, so that both clouds can be
seen on the same graph. The transverse profiles, which were
integrated out to arrive at the 1D model in the first place, are
recovered here for a 3D visualization.
\bibitem{Figure2-comment} Assuming a $^{87}$Rb$_{2}$ experiment, these
parameters coorespond to the molecular trapping potential with the axial
frequency $\omega _{x}/2\pi =0.01$ Hz (giving the harmonic oscillator length 
$l_{x}=\sqrt{\hbar /m_{2}\omega _{x}}=76.5$ $\mu $m, which we choose to be
our length scale, $d_{0}=l_{x}$), radial frequency $\omega _{\perp }/2\pi =82
$ Hz ($l_{\perp }=0.84$ $\mu $m), a 3D value of the atom-molecule coupling 
$\chi ^{(3D)}=4.95\times 10^{-7}$ m$^{3/2}$ s$^{-1}$ (so that the 1D
coupling is $\chi \simeq \chi ^{(3D)}/\sqrt{4\pi l_{\perp }^{2}}=0.165$ m$%
^{1/2}$ s$^{-1}$), and a molecule-molecule 3D scattering length $%
a_{22}=0.318$ nm ($U_{22}\simeq U_{22}^{(3D)}/(2\pi l_{\perp }^{2})=4\pi
\hbar a_{22}/(2\pi m_{2}l_{\perp }^{2})=3.28\times 10^{-7}$ m s$^{-1}$). The
relatively small value of $a_{22}$ is chosen here for making the computation
feasible; it corresponds to $a_{22}=0.06a_{11}$, where $a_{11}=5.3$ nm is
the scattering length for $^{87}$Rb atoms.
\end{thebibliography}
\end{document}